\documentclass[submission,copyright,creativecommons]{eptcs}
\providecommand{\event}{FMAS 2026} 

\usepackage{iftex}

\ifpdf
  \usepackage{underscore}         
  \usepackage[T1]{fontenc}        
\else
  \usepackage{breakurl}           
\fi

\title{Runtime Verification under Split Past and Future}
\author{Doğan Ulus
\institute{Department of Computer Engineering\\Boğaziçi University\\Istanbul, Turkey}
\email{dogan.ulus@bogazici.edu.tr}
}
\def\titlerunning{Runtime Verification under Split Past and Future}
\def\authorrunning{Doğan Ulus}

\usepackage{tikz}
\usetikzlibrary{
    arrows,
    arrows.meta,
    calc,
    intersections,
    shapes,
    shapes.arrows,
    shapes.geometric,
    patterns, 
    positioning, 
    decorations.markings,
    decorations.pathreplacing,
    decorations.text,
}

\usepackage{amsmath}
\usepackage{amssymb}
\usepackage{mathtools}
\usepackage{xspace}
\usepackage{lipsum}

\newcommand{\true}[0]{{\mathtt{true}}}
\newcommand{\false}[0]{{\mathtt{false}}}

\newcommand{\LTL}[0]{\textsf{LTL}\xspace}

\newcommand{\PastLTL}[0]{\textsf{PastLTL}\xspace}

\newcommand{\PROP}[0]{\textsf{AP}\xspace}
\newcommand{\FutureLTL}[0]{\textsf{FutureLTL}\xspace}
\newcommand{\SplitLTL}[0]{\textsf{SplitLTL}\xspace}

\DeclareMathOperator{\Until}{\texttt{U}}
\DeclareMathOperator{\Eventually}{\texttt{F}}
\DeclareMathOperator{\Always}{\texttt{G}}
\DeclareMathOperator*{\Next}{\texttt{X}}

\DeclareMathOperator{\WeakNext}{\texttt{WX}}
\DeclareMathOperator{\WeakUntil}{\texttt{WU}}

\DeclareMathOperator{\Since}{\texttt{S}}
\DeclareMathOperator{\Once}{\texttt{P}}
\DeclareMathOperator{\Hist}{\texttt{H}}
\DeclareMathOperator*{\Prev}{\texttt{Y}}

\DeclareMathOperator{\WeakPrev}{\texttt{WY}}
\DeclareMathOperator{\WeakSince}{\texttt{WS}}

\begin{document}
\maketitle

\begin{abstract}
Runtime assurance for autonomous systems increasingly requires reasoning not only about observed executions but also about anticipated future behaviors. 
Traditional runtime verification, however, primarily evaluates the execution observed so far and does not directly account for predicted continuations. We propose a formal framework for runtime assurance that combines monitoring of the observed execution with analysis of multiple predicted continuations. 
To support this integration, we introduce \emph{Split Linear Temporal Logic} (\SplitLTL), a linear-time temporal logic that assigns complementary roles to past and future temporal specifications within a single specification. 
The past component uses the observed history to determine the assurance requirements applicable at the current point of execution, while the future component evaluates predicted continuations against those requirements. 
The framework therefore filters predicted continuations according to the requirements induced by the observed history, identifying admissible continuations that can support subsequent decision making. 
We formally define the syntax and semantics of \SplitLTL and present an online monitoring architecture for evaluating observed executions together with predicted continuations.
\end{abstract}

\section{Introduction}
\label{sec:introduction}

Autonomous systems must make safety-critical decisions while operating in environments whose future evolution is only partially known. 
At any point during execution, an autonomous system has a concrete history of observations, but its future behavior remains uncertain and can only be estimated through prediction. 
Modern autonomous systems therefore routinely combine observations with predicted trajectories, behaviors, or system responses when making decisions. 
Ensuring these decisions are safe requires runtime reasoning over the observed history together with multiple possible future continuations.

Runtime Verification (RV) provides a natural mechanism for monitoring an execution as it unfolds. 
A runtime monitor incrementally observes the execution and evaluates temporal specifications against the behavior observed so far.
This retrospective perspective naturally favors past-time specifications, whose truth depends only on the observations available at the current monitoring point. 
Accordingly, several works have developed past-time approaches to runtime verification~\cite{HavelundR04,HavelundPU20,Ulus26Online}, motivated in part by the efficiency of online monitoring algorithms and the relative simplicity of their monitor constructions. 
Nevertheless, future temporal modalities remain prevalent in many RV frameworks, reflecting the influence of the future-oriented specification paradigm in formal verification.
In reactive settings, however, future modalities need not refer to an already determined future.
They can instead express requirements that constrain how the execution should proceed, a perspective first advocated by Gabbay~\cite{Gabbay87} and further developed in the \textsc{MetateM} executable temporal logic framework~\cite{BarringerFGO89}.

This paper similarly treats future temporal modalities as referring to possible future behavior, but places them in the context of runtime assurance with externally supplied predictions for autonomous decision making.
Rather than using temporal specifications as executable reactive programs, we use future modalities to express assurance obligations over predicted future continuations. 
To this end, we introduce \emph{Split Linear Temporal Logic} (\SplitLTL) in Section~\ref{sec:split-specs}, a~linear-time temporal logic that separates a~specification into past and future requirements evaluated relative to the current time point. 
The key idea is to evaluate the past component against the actual execution prefix and the future component against the predicted continuations available at that point.

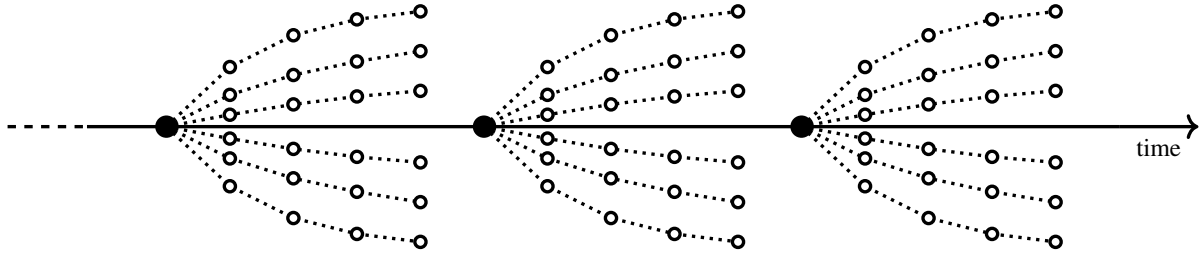
\begin{figure*}[t]
\centering
\resizebox{\textwidth}{!}{
\begin{tikzpicture}[line width=1.2pt]


\draw[dashed] (-2,0) -- (-1,0);
\draw (-1,0) -- (12,0);
\draw[->] (12,0) -- (13,0);

\foreach \x in {0,4,8}
    \fill (\x,0) circle (4pt);


\draw[dotted]
    (0,0) -- (0.8,0.75) -- (1.6,1.15) -- (2.4,1.35) -- (3.2,1.45);
\foreach \p in {(0.8,0.75),(1.6,1.15),(2.4,1.35),(3.2,1.45)}
    \filldraw[fill=white] \p circle (2pt);

\draw[dotted]
    (0,0) -- (0.8,0.40) -- (1.6,0.65) -- (2.4,0.82) -- (3.2,0.95);
\foreach \p in {(0.8,0.40),(1.6,0.65),(2.4,0.82),(3.2,0.95)}
    \filldraw[fill=white] \p circle (2pt);

\draw[dotted]
    (0,0) -- (0.8,0.15) -- (1.6,0.28) -- (2.4,0.38) -- (3.2,0.45);
\foreach \p in {(0.8,0.15),(1.6,0.28),(2.4,0.38),(3.2,0.45)}
    \filldraw[fill=white] \p circle (2pt);

\draw[dotted]
    (0,0) -- (0.8,-0.15) -- (1.6,-0.28) -- (2.4,-0.38) -- (3.2,-0.45);
\foreach \p in {(0.8,-0.15),(1.6,-0.28),(2.4,-0.38),(3.2,-0.45)}
    \filldraw[fill=white] \p circle (2pt);

\draw[dotted]
    (0,0) -- (0.8,-0.40) -- (1.6,-0.65) -- (2.4,-0.82) -- (3.2,-0.95);
\foreach \p in {(0.8,-0.40),(1.6,-0.65),(2.4,-0.82),(3.2,-0.95)}
    \filldraw[fill=white] \p circle (2pt);

\draw[dotted]
    (0,0) -- (0.8,-0.75) -- (1.6,-1.15) -- (2.4,-1.35) -- (3.2,-1.45);
\foreach \p in {(0.8,-0.75),(1.6,-1.15),(2.4,-1.35),(3.2,-1.45)}
    \filldraw[fill=white] \p circle (2pt);


\draw[dotted]
    (4,0) -- (4.8,0.75) -- (5.6,1.15) -- (6.4,1.35) -- (7.2,1.45);
\foreach \p in {(4.8,0.75),(5.6,1.15),(6.4,1.35),(7.2,1.45)}
    \filldraw[fill=white] \p circle (2pt);

\draw[dotted]
    (4,0) -- (4.8,0.40) -- (5.6,0.65) -- (6.4,0.82) -- (7.2,0.95);
\foreach \p in {(4.8,0.40),(5.6,0.65),(6.4,0.82),(7.2,0.95)}
    \filldraw[fill=white] \p circle (2pt);

\draw[dotted]
    (4,0) -- (4.8,0.15) -- (5.6,0.28) -- (6.4,0.38) -- (7.2,0.45);
\foreach \p in {(4.8,0.15),(5.6,0.28),(6.4,0.38),(7.2,0.45)}
    \filldraw[fill=white] \p circle (2pt);

\draw[dotted]
    (4,0) -- (4.8,-0.15) -- (5.6,-0.28) -- (6.4,-0.38) -- (7.2,-0.45);
\foreach \p in {(4.8,-0.15),(5.6,-0.28),(6.4,-0.38),(7.2,-0.45)}
    \filldraw[fill=white] \p circle (2pt);

\draw[dotted]
    (4,0) -- (4.8,-0.40) -- (5.6,-0.65) -- (6.4,-0.82) -- (7.2,-0.95);
\foreach \p in {(4.8,-0.40),(5.6,-0.65),(6.4,-0.82),(7.2,-0.95)}
    \filldraw[fill=white] \p circle (2pt);

\draw[dotted]
    (4,0) -- (4.8,-0.75) -- (5.6,-1.15) -- (6.4,-1.35) -- (7.2,-1.45);
\foreach \p in {(4.8,-0.75),(5.6,-1.15),(6.4,-1.35),(7.2,-1.45)}
    \filldraw[fill=white] \p circle (2pt);


\draw[dotted]
    (8,0) -- (8.8,0.75) -- (9.6,1.15) -- (10.4,1.35) -- (11.2,1.45);
\foreach \p in {(8.8,0.75),(9.6,1.15),(10.4,1.35),(11.2,1.45)}
    \filldraw[fill=white] \p circle (2pt);

\draw[dotted]
    (8,0) -- (8.8,0.40) -- (9.6,0.65) -- (10.4,0.82) -- (11.2,0.95);
\foreach \p in {(8.8,0.40),(9.6,0.65),(10.4,0.82),(11.2,0.95)}
    \filldraw[fill=white] \p circle (2pt);

\draw[dotted]
    (8,0) -- (8.8,0.15) -- (9.6,0.28) -- (10.4,0.38) -- (11.2,0.45);
\foreach \p in {(8.8,0.15),(9.6,0.28),(10.4,0.38),(11.2,0.45)}
    \filldraw[fill=white] \p circle (2pt);

\draw[dotted]
    (8,0) -- (8.8,-0.15) -- (9.6,-0.28) -- (10.4,-0.38) -- (11.2,-0.45);
\foreach \p in {(8.8,-0.15),(9.6,-0.28),(10.4,-0.38),(11.2,-0.45)}
    \filldraw[fill=white] \p circle (2pt);

\draw[dotted]
    (8,0) -- (8.8,-0.40) -- (9.6,-0.65) -- (10.4,-0.82) -- (11.2,-0.95);
\foreach \p in {(8.8,-0.40),(9.6,-0.65),(10.4,-0.82),(11.2,-0.95)}
    \filldraw[fill=white] \p circle (2pt);

\draw[dotted]
    (8,0) -- (8.8,-0.75) -- (9.6,-1.15) -- (10.4,-1.35) -- (11.2,-1.45);
\foreach \p in {(8.8,-0.75),(9.6,-1.15),(10.4,-1.35),(11.2,-1.45)}
    \filldraw[fill=white] \p circle (2pt);



\node[below] at (12.5,0) {\footnotesize time};

\end{tikzpicture}
}
\caption{Multiple predicted future states generated at successive time points.}

\label{fig:intro}
\end{figure*}

We outline our \SplitLTL monitoring framework in Section~\ref{sec:framework}. 
At each time point, we assume the system under monitoring supplies one or more predicted trajectories describing its future states and environment over a finite prediction horizon, as illustrated in Figure~\ref{fig:intro}. 
The monitor evaluates each predicted trajectory against the future-time requirements for those rules whose corresponding past-time requirements are satisfied by the actual execution prefix.
Predicted trajectories that violate the requirements are thereby identified as inadmissible, while the remaining trajectories constitute the set of continuations consistent with the specified assurance requirements. 
The decision-making mechanism can then use this filtered set of predicted trajectories when selecting an action at the current time point.

\section{Preliminaries}

Linear Temporal Logic (\LTL) extends the propositional logic with past and future temporal modalities such as \emph{Previously} ($\Prev$), \emph{Next} ($\Next$), \emph{Since} ($\Since$) and \emph{Until} ($\Until$) modalities.
Given a finite set $\PROP$ of atomic predicates, \LTL formulas are inductively defined by the grammar
\begin{equation*}
\varphi\ \coloneqq\ \true\ |\  \false\ |\ p \ |\ \neg \varphi\ |\ \varphi_{1} \wedge \varphi_{2}\  |\ \Prev\varphi\ |\ \Next\varphi\ |\ \varphi_{1} \Since \varphi_{2}\ |\ \varphi_{1} \Until \varphi_{2}
\end{equation*}
where $p\in\PROP$. We distinguish two syntactic fragments of \LTL. The past fragment $\PastLTL$ consists of formulas generated by the grammar obtained by excluding $\Next$ and $\Until$; analogously, the future fragment $\FutureLTL$ is obtained by excluding $\Prev$ and $\Since$.

A (possibly finite or infinite) discrete-time behavior is a function $w:\mathbb{T}\to\mathbb{B}^{\PROP}$, where $\mathbb{T}\subseteq\mathbb{N}$ is a non-empty interval of discrete time points and $\mathbb{B}=\{\true,\false\}$ is the Boolean domain.
The satisfaction of an \LTL formula $\varphi$ at a given time instant $t \in\mathbb{T}$ over a discrete-time behavior $w$, denoted as $(w, t) \vDash \varphi$, is defined inductively.
We first define the semantics of the propositional fragment of \LTL:

\begin{equation*}
\setlength{\arraycolsep}{2pt}
\begin{array}{rlcl}
(w,t) &\vDash \true & \ \Longleftrightarrow\ & \true\\
(w,t) &\vDash \false & \ \Longleftrightarrow\ & \false\\
(w,t) &\vDash p & \ \Longleftrightarrow\ & w_{p}(t) = \true  \\
(w,t) &\vDash \neg \varphi & \ \Longleftrightarrow\ & (w,t) \nvDash \varphi \\
(w,t) &\vDash \varphi_{1} \wedge \varphi_{2} & \ \Longleftrightarrow\ & (w,t) \vDash \varphi_{1} \land (w, t) \vDash \varphi_{2}
\end{array}
\label{DefProp}
\end{equation*}
The remaining Boolean operators, including disjunction ($\vee$), logical implication ($\rightarrow$), and biconditional ($\leftrightarrow$), are defined as usual in terms of negation ($\neg$) and conjunction ($\wedge$).
For the temporal modalities \emph{Previously} ($\Prev$) and
\emph{Next} ($\Next$), the semantics is defined with respect to the
immediately preceding and succeeding time instants, respectively:
\begin{equation*}
\setlength{\arraycolsep}{2pt}
\begin{array}{rlcl}
(w,t) &\vDash \Prev\varphi &\ \Longleftrightarrow\ &t-1 \in\mathbb{T}\, \land (w,\,t - 1)\vDash\varphi\\
(w,t) &\vDash \Next\varphi &\ \Longleftrightarrow\ &t+1 \in\mathbb{T}\, \land (w,\,t + 1)\vDash\varphi\\
\end{array}
\label{DefPrevNext}
\end{equation*}
For the temporal modalities \emph{Since} ($\Since$) and \emph{Until} ($\Until$), we adopt reflexive semantics, where the current time instant may serve as the witness. The semantics of these modalities are defined, respectively, as follows:
\begin{equation*}
\setlength{\arraycolsep}{2pt}
\begin{array}{rlcl}
(w,t) &\vDash \varphi_{1}\Since\varphi_{2}
&\ \Longleftrightarrow\ &
\exists t'\in\mathbb{T}.
\bigl(
t'\leq t
\land
(w,t')\vDash\varphi_{2}
\land
\forall t''\in\mathbb{T}.
\bigl(t'< t'' \leq t
\Rightarrow
(w,t'')\vDash\varphi_{1}\bigr)
\bigr)\\[1mm]
(w,t) &\vDash \varphi_{1}\Until\varphi_{2}
&\ \Longleftrightarrow\ &
\exists t'\in\mathbb{T}.
\bigl(
t \leq t'
\land
(w,t')\vDash\varphi_{2}
\land
\forall t''\in\mathbb{T}.
\bigl(t \leq t'' < t'
\Rightarrow
(w,t'')\vDash\varphi_{1}\bigr)
\bigr)
\end{array}
\label{DefS}
\end{equation*}
The following commonly used temporal modalities are treated as syntactic abbreviations: 
\emph{Sometime in the Past} ($\Once$), 
\emph{Always in the Past} ($\Hist$), 
\emph{Weak Previous} ($\WeakPrev$)
and \emph{Weak Since} ($\WeakSince$) 
for the past-time fragment $\PastLTL$, 
and \emph{Eventually} ($\Eventually$), 
\emph{Always} ($\Always$), 
\emph{Weak Next} ($\WeakNext$)
and \emph{Weak Until} ($\WeakUntil$) 
for the future-time fragment $\FutureLTL$:
\begin{equation*}
\begin{aligned}
\Once\varphi &\equiv \true\Since\varphi &\quad&\quad&\Eventually\varphi &\equiv \true\Until\varphi\\
\Hist\varphi &\equiv \neg\Once\neg\varphi & &&\Always\varphi &\equiv \neg\Eventually\neg\varphi\\
\WeakPrev\varphi &\equiv \neg\Prev\true \vee \Prev\varphi & &&\WeakNext\varphi &\equiv \neg\Next\true \vee \Next\varphi\\
\varphi_1\WeakSince\varphi_2 &\equiv (\varphi_1\Since\varphi_2)\vee\Hist\varphi_1&&&\varphi_1\WeakUntil\varphi_2&\equiv (\varphi_1\Until\varphi_2)\vee\Always\varphi_1\\
\end{aligned}
\label{EqLTL}
\end{equation*}
The weak temporal operators are particularly useful for finite behaviors, as they provide well-defined semantics when the observed or predicted behavior is too short.

\section{Split Temporal Specifications}
\label{sec:split-specs}

In this section, we introduce Split Linear Temporal Logic (\SplitLTL), which provides a formal structure for associating a past-time condition (antecedent) with a future-time requirement (consequent).
We start by defining a \SplitLTL rule~$\phi$ as a pair of formulas such that
$$\phi \coloneqq (\varphi_{\textsf{past}},\, \varphi_{\textsf{future}})$$
where $\varphi_{\textsf{past}} \in \PastLTL$ and $\varphi_{\textsf{future}} \in \FutureLTL$ denote the past and future components of the rule~$\phi$, respectively. We then define \SplitLTL formulas by the grammar
\begin{equation*}
\Phi\ \coloneqq\ \true\ |\  \false\ |\ \phi \ |\ \neg \Phi\ |\ \Phi_{1} \wedge \Phi_{2}
\end{equation*}
where $\phi$ denotes a \SplitLTL rule. 
The satisfaction of a \SplitLTL rule $\phi = (\varphi_{\textsf{past}},\, \varphi_{\textsf{future}})$ at a given time instant $t \in\mathbb{T}$ over a discrete-time behavior $w$ is defined by
\begin{equation*}
\setlength{\arraycolsep}{2pt}
\begin{array}{rlcl}
(w,t) &\vDash \phi &\ \Longleftrightarrow\ & (w,t) \nvDash \varphi_{\textsf{past}} \vee (w,t) \vDash \varphi_{\textsf{future}}\\
\end{array}
\label{DefSplitRule}
\end{equation*}
This formulation is semantically equivalent to the \LTL formula $\varphi_{\textsf{past}}\rightarrow\varphi_{\textsf{future}}$. This syntactic form is inspired by prior work on declarative temporal logic and executable temporal specifications~\cite{Gabbay87, BarringerFGO89} and is grounded in Gabbay's Separation Theorem~\cite{Gabbay81Separation}. However, rather than deriving the separation through a transformation of an arbitrary \LTL formula, \SplitLTL enforces the separation explicitly at the syntactic level by requiring each rule to consist of a past-time component and a future-time component. 

Finally, we must note that the classical result stating that past-time modalities do not extend the expressive power of \FutureLTL assumes evaluation at a fixed initial time point $t=0$. Although this assumption is suitable for model checking settings, we employ a moving evaluation point in our monitoring framework. 
Thus, retaining both past- and future-time modalities is natural for runtime verification and provides the basis for the separation used by \SplitLTL.

\section{Split Temporal Logic Monitoring Architecture}
\label{sec:framework}

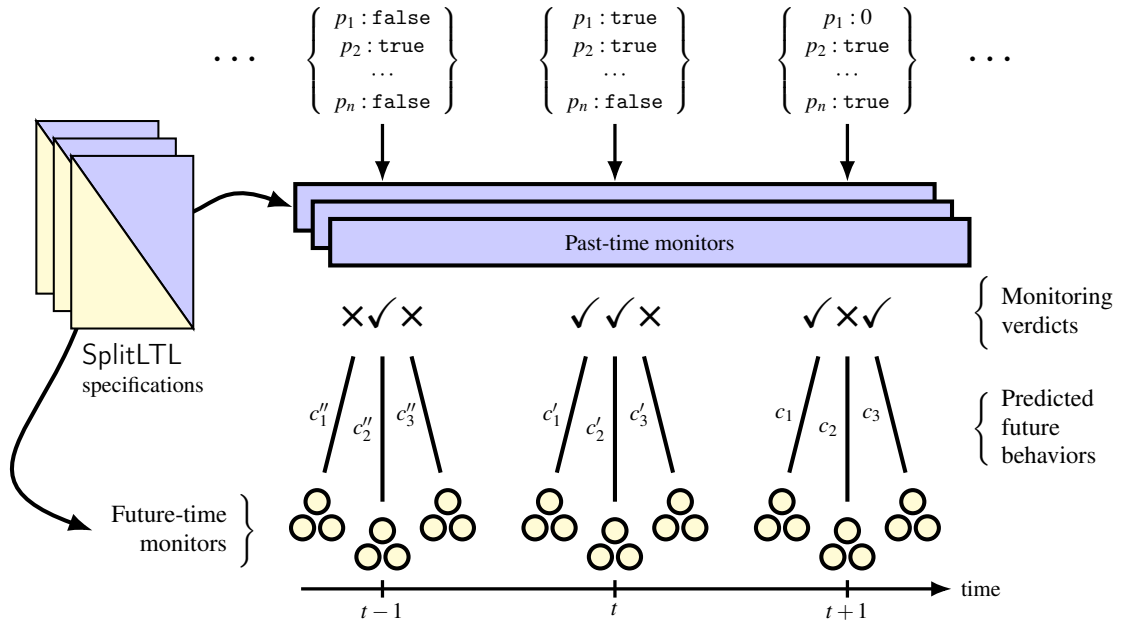
\begin{figure}[b!]
\centering
\resizebox{\columnwidth}{!}{%
\begin{tikzpicture}[
    >=Latex,
    box/.style={draw, fill=blue!20, minimum width=11cm, minimum height=0.8cm, align=center, line width=2pt},
    smallbox/.style={draw, fill=yellow!20, minimum width=2.10cm, minimum height=2.97cm, align=center, line width=2pt},
    vec/.style={align=center},
    arr/.style={-Latex, ultra thick},
    compilearrow/.style={draw, fill=blue!20, single arrow, single arrow head extend=2mm, minimum height=1cm, minimum width=1.6cm}
]

\node[] (x1front) at (0,0) {X1};
\node[box, right=3cm of x1front] (monitor3) {};
\node[box, behind path, xshift=0.3cm, yshift=-0.3cm] (monitor2) at (monitor3.center) {};
\node[box, behind path, xshift=0.3cm, yshift=-0.3cm] (monitor) at (monitor2.center) {Past-time monitors};

\node[vec, above=1cm of monitor3, xshift=-4cm] (p1) {$\left\{\begin{array}{c}p_1: \false\\p_2: \true\\\cdots\\p_n: \false\end{array}\right\}$};
\node[vec, above=1cm of monitor3] (p2) {$\left\{\begin{array}{c}p_1: \true\\p_2: \true\\\cdots\\p_n: \false\end{array}\right\}$};
\node[vec, above=1cm of monitor3, xshift=4cm] (p3) {$\left\{\begin{array}{c}p_1: 0\\p_2: \true\\\cdots\\p_n: \true\end{array}\right\}$};

\node[align=center] (fmon) at ($(p1.south)+(-3.5,-7)$) 
    {\large $\begin{array}{r} \text{Future-time} \\ \text{monitors} \end{array}\Biggr\}$};

\draw[line width=2pt] ($(p1.south)+(-0.5,-4)$) -- ++(-0.5,-2) node[midway, left] {$c''_{1}$};
\draw[line width=2pt] ($(p1.south)+(0,-4)$) -- ++(0,-2.5) node[midway, left] {$c''_{2}$};
\draw[line width=2pt] ($(p1.south)+(0.5,-4)$) -- ++(0.5,-2) node[midway, left] {$c''_{3}$};

\node[] (p1tick) at ($(p1.south)+(0,-3.25)$) {\huge \texttimes\checkmark\texttimes};

\node[circle,fill=yellow!20,inner sep=4pt,line width=2pt,draw] (p1p11) at ($(p1.south)+(-1.125,-6.5)$) {};
\node[circle,fill=yellow!20,inner sep=4pt,line width=2pt,draw] (p1p12) at ($(p1p11)+(-0.25,-0.45)$) {};
\node[circle,fill=yellow!20,inner sep=4pt,line width=2pt,draw] (p1p13) at ($(p1p11)+(0.25,-0.45)$) {};
\node[circle,fill=yellow!20,inner sep=4pt,line width=2pt,draw] (p1p21) at ($(p1.south)+(0,-7)$) {};
\node[circle,fill=yellow!20,inner sep=4pt,line width=2pt,draw] (p1p22) at ($(p1p21)+(-0.25,-0.45)$) {};
\node[circle,fill=yellow!20,inner sep=4pt,line width=2pt,draw] (p1p23) at ($(p1p21)+(0.25,-0.45)$) {};
\node[circle,fill=yellow!20,inner sep=4pt,line width=2pt,draw] (p1p31) at ($(p1.south)+(1.125,-6.5)$) {};
\node[circle,fill=yellow!20,inner sep=4pt,line width=2pt,draw] (p1p32) at ($(p1p31)+(-0.25,-0.45)$) {};
\node[circle,fill=yellow!20,inner sep=4pt,line width=2pt,draw] (p1p33) at ($(p1p31)+(0.25,-0.45)$) {};

\draw[line width=2pt] ($(p2.south)+(-0.5,-4)$) -- ++(-0.5,-2) node[midway, left] {$c'_{1}$};
\draw[line width=2pt] ($(p2.south)+(0,-4)$) -- ++(0,-2.5) node[midway, left] {$c'_{2}$};
\draw[line width=2pt] ($(p2.south)+(0.5,-4)$) -- ++(0.5,-2) node[midway, left] {$c'_{3}$};

\node[] (p2tick) at ($(p2.south)+(0,-3.25)$) {\huge \checkmark\checkmark\texttimes};

\node[circle,fill=yellow!20,inner sep=4pt,line width=2pt,draw] (p2p11) at ($(p2.south)+(-1.125,-6.5)$) {};
\node[circle,fill=yellow!20,inner sep=4pt,line width=2pt,draw] (p2p12) at ($(p2p11)+(-0.25,-0.45)$) {};
\node[circle,fill=yellow!20,inner sep=4pt,line width=2pt,draw] (p2p13) at ($(p2p11)+(0.25,-0.45)$) {};
\node[circle,fill=yellow!20,inner sep=4pt,line width=2pt,draw] (p2p21) at ($(p2.south)+(0,-7)$) {};
\node[circle,fill=yellow!20,inner sep=4pt,line width=2pt,draw] (p2p22) at ($(p2p21)+(-0.25,-0.45)$) {};
\node[circle,fill=yellow!20,inner sep=4pt,line width=2pt,draw] (p2p23) at ($(p2p21)+(0.25,-0.45)$) {};
\node[circle,fill=yellow!20,inner sep=4pt,line width=2pt,draw] (p2p31) at ($(p2.south)+(1.125,-6.5)$) {};
\node[circle,fill=yellow!20,inner sep=4pt,line width=2pt,draw] (p2p32) at ($(p2p31)+(-0.25,-0.45)$) {};
\node[circle,fill=yellow!20,inner sep=4pt,line width=2pt,draw] (p2p33) at ($(p2p31)+(0.25,-0.45)$) {};

\draw[line width=2pt] ($(p3.south)+(-0.5,-4)$) -- ++(-0.5,-2) node[midway, left] {$c_{1}$};
\draw[line width=2pt] ($(p3.south)+(0,-4)$) -- ++(0,-2.5) node[midway, left] {$c_{2}$};
\draw[line width=2pt] ($(p3.south)+(0.5,-4)$) -- ++(0.5,-2) node[midway, left] {$c_{3}$};

\node[] (p3tick) at ($(p3.south)+(0,-3.25)$) {\huge \checkmark\texttimes\checkmark};

\node[circle,fill=yellow!20,inner sep=4pt,line width=2pt,draw] (p3p11) at ($(p3.south)+(-1.125,-6.5)$) {};
\node[circle,fill=yellow!20,inner sep=4pt,line width=2pt,draw] (p3p12) at ($(p3p11)+(-0.25,-0.45)$) {};
\node[circle,fill=yellow!20,inner sep=4pt,line width=2pt,draw] (p3p13) at ($(p3p11)+(0.25,-0.45)$) {};
\node[circle,fill=yellow!20,inner sep=4pt,line width=2pt,draw] (p3p21) at ($(p3.south)+(0,-7)$) {};
\node[circle,fill=yellow!20,inner sep=4pt,line width=2pt,draw] (p3p22) at ($(p3p21)+(-0.25,-0.45)$) {};
\node[circle,fill=yellow!20,inner sep=4pt,line width=2pt,draw] (p3p23) at ($(p3p21)+(0.25,-0.45)$) {};
\node[circle,fill=yellow!20,inner sep=4pt,line width=2pt,draw] (p3p31) at ($(p3.south)+(1.125,-6.5)$) {};
\node[circle,fill=yellow!20,inner sep=4pt,line width=2pt,draw] (p3p32) at ($(p3p31)+(-0.25,-0.45)$) {};
\node[circle,fill=yellow!20,inner sep=4pt,line width=2pt,draw] (p3p33) at ($(p3p31)+(0.25,-0.45)$) {};

\draw[line width=2pt,->] (x1front.east) to[out=-20, in=160, looseness=1.5] (monitor3.west);
\draw[line width=2pt,->] (x1front.south) to[out=-90, in=160, looseness=1.5] (fmon.west);

\node[align=left] (assess) at ($(p3.south)+(2,-3.25cm)$) [right]
    {\large $\Bigg\{ \begin{array}{l} \text{Monitoring}\\\text{verdicts} \end{array}$};
\node[align=left] (assess) at ($(p3.south)+(2,-5.25cm)$) [right]
    {\large $\Bigg\{ \begin{array}{l} \text{Predicted}\\\text{future} \\\text{behaviors} \end{array}$};

\node[below=1cm of monitor, xshift=-4cm] (y1) {};
\node[below=1cm of monitor] (y2) {};
\node[below=1cm of monitor, xshift=4cm] (y3) {};

\node[scale=2] at ($(p1)-(2.5,0)$) {$\cdots$};
\node[scale=2] at ($(p3)+(2.5,0)$) {$\cdots$};

\draw[arr] (p1.south) -- ++(0,-0.6) -- ($(monitor3.north)+(-4cm,0)$);
\draw[arr] (p2.south) -- ++(0,-0.6) -- ($(monitor3.north)$);
\draw[arr] (p3.south) -- ++(0,-0.6) -- ($(monitor3.north)+(+4cm,0)$);


\draw[-Latex, ultra thick] ($(p1.south)+(-1.4,-8)$) -- ($(p3.south)+(1.8,-8)$)
    node[right] {time};

\draw[ultra thick] ($(p1.south)+(0,-8.15)$) -- ($(p1.south)+(0,-7.85)$);
\draw[ultra thick] ($(p2.south)+(0,-8.15)$) -- ($(p2.south)+(0,-7.85)$);
\draw[ultra thick] ($(p3.south)+(0,-8.15)$) -- ($(p3.south)+(0,-7.85)$);

\node[below=3pt] at ($(p1.south)+(0,-8)$) {$t-1$};
\node[below=3pt] at ($(p2.south)+(0,-8)$) {$t$};
\node[below=3pt] at ($(p3.south)+(0,-8)$) {$t+1$};


\path[
    draw,
    line width=1pt,
    fill=blue!20
]
    ($(x1front.center)+(-1.05cm,1.485cm)$) --
    ($(x1front.center)+(1.05cm,1.485cm)$) --
    ($(x1front.center)+(1.05cm,-1.485cm)$) --
    cycle;

\path[
    draw,
    line width=1pt,
    fill=yellow!20
]
    ($(x1front.center)+(-1.05cm,1.485cm)$) --
    ($(x1front.center)+(-1.05cm,-1.485cm)$) --
    ($(x1front.center)+(1.05cm,-1.485cm)$) --
    cycle;

\path[
    draw,
    line width=1pt,
    fill=blue!20
]
    ($(x1front.center)+(-0.75cm,1.185cm)$) --
    ($(x1front.center)+(1.35cm,1.185cm)$) --
    ($(x1front.center)+(1.35cm,-1.785cm)$) --
    cycle;

\path[
    draw,
    line width=1pt,
    fill=yellow!20
]
    ($(x1front.center)+(-0.75cm,1.185cm)$) --
    ($(x1front.center)+(-0.75cm,-1.785cm)$) --
    ($(x1front.center)+(1.35cm,-1.785cm)$) --
    cycle;

\path[
    draw,
    line width=1pt,
    fill=blue!20
]
    ($(x1front.center)+(-0.45cm,0.885cm)$) --
    ($(x1front.center)+(1.65cm,0.885cm)$) --
    ($(x1front.center)+(1.65cm,-2.085cm)$) --
    cycle;

\path[
    draw,
    line width=1pt,
    fill=yellow!20
]
    ($(x1front.center)+(-0.45cm,0.885cm)$) --
    ($(x1front.center)+(-0.45cm,-2.085cm)$) --
    ($(x1front.center)+(1.65cm,-2.085cm)$) --
    cycle;
    
\node[align=left] at ($(x1front.center)+(-0.4cm,-2.2cm)$) [below right] {\Large \SplitLTL\\specifications};

\end{tikzpicture}%
}
\caption{Split temporal logic monitoring architecture.}
\label{fig:architecture}
\end{figure}

This section outlines our architecture for the \SplitLTL monitoring framework.
We consider autonomous systems equipped with a specific class of controllers~\cite{FoxBT97,WerlingZKT10,UlusB19} that, at each evaluation point, generate one or more predicted behaviors of their future states and the environment over a finite prediction horizon.
These controllers are often designed to filter and rank the predicted behaviors to select their next control action.
Our monitoring framework supports this decision process by evaluating formally specified requirements over the observed history and the set of predicted behaviors.
Figure~\ref{fig:architecture} depicts the overall monitoring architecture.

Given a \SplitLTL formula, our architecture consists of a separate past-time monitor for the antecedent and a future-time monitor for the consequent of each rule in the formula.
During runtime monitoring, the observed history is fed incrementally, one observation at a time, to all past-time monitors, which update their internal states and evaluate the corresponding past-time specifications without reprocessing the entire history. 
Further, at each time point, we assume that a set $\mathcal{C}_t$ of predicted future behaviors is provided to the monitoring framework. The future-time monitors evaluate each predicted behavior $c\in\mathcal{C}_t$ against the corresponding future-time requirement and produce a verdict for each behavior. 
These verdicts are then combined with the verdicts of the corresponding past-time monitors to determine whether the \SplitLTL formula is satisfied. 
Unlike the past-time monitors, which maintain state across observations of the shared history, the future-time monitors are reset for each predicted behavior at every time point. 
Thus, the computational cost of monitoring grows with the number and length of predicted behaviors. 
Since these predictions must be processed within the controller's control cycle, efficient future-time monitoring is critical for real-time operation.

We implement our framework using Reelay~\cite{ulus2026reelay}, an online runtime verification library for past-time temporal logic specifications, including \PastLTL. 
Reelay employs sequential network-based past-time monitors for efficient online monitoring~\cite{Ulus26Online}, processing each observation in time linear in the size of the monitored formula. 
Empirical measurements report per-observation latencies in the range of tens to hundreds of nanoseconds, depending on the message transport and message size. 
We have also shown that sharing common subformulas and underlying data structures further reduces monitoring overhead when monitoring multiple formulas~\cite{demir2026multi}. 
In our \SplitLTL monitoring architecture, we construct Reelay monitors directly from the past-time component of each rule and share monitors for common subformulas across rules whenever possible. 
These capabilities provide an efficient foundation for the past-time monitoring component of our framework.

\begin{figure*}[t]
\centering
\resizebox{\textwidth}{!}{
\begin{tikzpicture}[line width=1.2pt]

\draw[dashed] (1,0) -- (2,0);
\draw (2,0) to[out=0, in=180]      
    node[solid, pos=0, circle, draw, fill, inner sep=1.5pt]{}
    node[pos=0, below] {\scriptsize$t-2$}
    node[solid, pos=0.5, circle, draw, fill, inner sep=1.5pt]{}
    node[pos=0.5, below] {\scriptsize$t-1$}
    node[solid, pos=1, circle, draw, fill, inner sep=1.5pt]{}
    node[pos=1, below] {\scriptsize$t$}
    (6,0);
\draw[color=black!40, ->] (4,0.6) -- (5.5,0.6);

\draw[dotted] (6,0) to[out=0, in=180]
    node[solid, pos=0.33, circle, draw, inner sep=1.5pt]{}
    node[pos=0.33, below] {\scriptsize$t+1$}
    node[solid, pos=0.67, circle, draw, inner sep=1.5pt]{}
    node[pos=0.67, below] {\scriptsize$t+2$}
    node[solid, pos=1, circle, draw, inner sep=1.5pt]{}
    node[pos=1, below] {\scriptsize$t+3$}
(12,1.5) -- (13,1.5);

\draw[dotted] (6,0) to[out=0, in=180] 
    node[solid, pos=0.33, circle, draw, inner sep=1.5pt]{}
    node[pos=0.33, below] {\scriptsize$t+1$}
    node[solid, pos=0.67, circle, draw, inner sep=1.5pt]{}
    node[pos=0.67, below] {\scriptsize$t+2$}
    node[solid, pos=1, circle, draw, inner sep=1.5pt]{}
    node[pos=1, below] {\scriptsize$t+3$}
(12,-1.5) -- (13,-1.5);

\path[
  postaction={
    decorate,
    decoration={
      markings,
      mark=at position 0.5 with {\coordinate (P);}
    }
  }
]
(12,1.5) to[out=180,in=0] (6,0);

\begin{scope}[yshift=0.5cm]
  \clip (P |- -1,0) rectangle (13,3);

  \draw[
    color=black!40,
    postaction={
      decorate,
      decoration={
        markings,
        mark=at position 0.5 with {\arrow{>}}
      }
    }
  ]
  (12,1.5) to[out=180,in=0] (6,0);
\end{scope}

\path[
  postaction={
    decorate,
    decoration={
      markings,
      mark=at position 0.5 with {\coordinate (P);}
    }
  }
]
(12,-1.5) to[out=180,in=0] (6,0);

\begin{scope}[yshift=0.5cm]
  \clip (P |- -1,0) rectangle (13,-3);

  \draw[
    color=black!40,
    postaction={
      decorate,
      decoration={
        markings,
        mark=at position 0.5 with {\arrow{>}}
      }
    }
  ]
  (12,-1.5) to[out=180,in=0] (6,0);
\end{scope}

\node[rectangle,fill=blue!20,inner sep=8pt,line width=2pt,draw] (m2) at (6,0.6) {};
\node[circle,fill=yellow!20,inner sep=4pt,line width=2pt,draw] (m1) at (12,2) {};
\node[circle,fill=yellow!20,inner sep=4pt,line width=2pt,draw] (m3) at (12,-1) {};

\node[above=6pt] (mtext2) at (12,2) {\scriptsize $\mathfrak{M}\bigl(\textsf{rev}(\varphi_{\textsf{future}})\bigr)$};
\node[above=6pt] (mtext2) at (12,-1) {\scriptsize $\mathfrak{M}\bigl(\textsf{rev}(\varphi_{\textsf{future}})\bigr)$};
\node[above=6pt] (mtext2) at (6,0.75) {\scriptsize $\mathfrak{M}(\varphi_{\textsf{past}})$};

\node[right] (mtext2) at (1,2) {\scriptsize $\phi\coloneqq(\varphi_{\textsf{past}},\,\varphi_{\textsf{future}})$};

\end{tikzpicture}
}
\caption{Evaluation of an individual \SplitLTL rule $\phi\coloneqq(\varphi_{\textsf{past}},\,\varphi_{\textsf{future}})$ over multiple predicted behaviors by running past-time monitors from both ends.}
\label{fig:split-behavior}
\end{figure*}
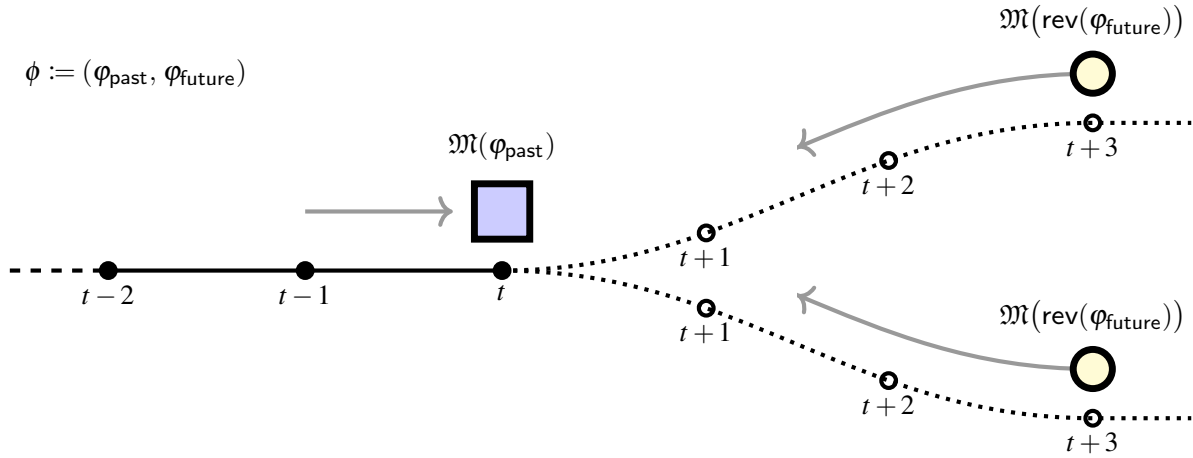

For the future-time monitors, we exploit the temporal symmetry between the past-time and future-time fragments of \LTL. 
Specifically, a future-time formula $\varphi_{\textsf{future}}$ can be transformed into a corresponding past-time formula $\varphi'{\textsf{past}} = \textsf{rev}(\varphi{\textsf{future}})$ by reversing the temporal direction and replacing future-time operators with their past-time counterparts. 
Given a predicted behavior, the past-time monitor $\mathfrak{M}(\textsf{rev}(\varphi_{\textsf{future}}))$ constructed from the reversed formula processes its states in reverse temporal order, starting from the last state and proceeding to the current time point.
The monitor's verdict on the reversed behavior matches that of the original future-time specification on the predicted behavior. 
Figure~\ref{fig:split-behavior} illustrates this two-sided monitoring technique, in which both the observed history and predicted behaviors are evaluated using past-time monitors. 
This symmetry allows us to reuse the same high-performance monitoring infrastructure for both past- and future-time formulas and eliminates the need for a separate future-time monitoring engine.

Finally, although prediction and selection modules play a critical role in the overall decision-making process, our monitoring framework remains independent of how predictions are generated and selected. Our monitoring architecture only requires the prediction mechanism to provide individual finite state sequences, which may be obtained through forward simulation under control inputs, learning-enabled techniques, or probabilistic models. This allows the framework to provide runtime assurance across diverse prediction and control architectures by treating predicted behaviors as inputs to the monitoring process.



\section{Related Work}
\label{sec:related}

Several works have extended runtime verification by predicting the satisfaction or violation of a property from a temporal behavior. Uncertainty about future observations has motivated approaches based on three-valued~\cite{ZhangLD12} and probabilistic semantics~\cite{QinD20}, as well as the orchestration of multiple prediction models~\cite{FerrandoCFLPFM21}. In contrast, \SplitLTL evaluates predicted system behaviors to support autonomous decision-making rather than to derive early verdicts on system correctness. Its role is closer to candidate evaluation in search-based planning~\cite{WerlingZKT10}, where candidate future behaviors are assessed against history-dependent temporal requirements encoded as formal monitors.

Shield synthesis and runtime assurance address a complementary problem: ensuring that system execution respects safety constraints. Shield synthesis constructs runtime enforcement mechanisms that intervene when a system's proposed actions may violate a formal specification~\cite{KonighoferABHKT17}. Runtime assurance architectures, in turn, combine an advanced controller with a safety mechanism that can intervene when the controller's behavior threatens system safety. Such mechanisms include safety controllers synthesized using backward reachability analysis~\cite{SchurmannKKA22}. The proposed \SplitLTL monitoring architecture instead employs multiple efficient formal monitors to filter candidate actions according to their predicted consequences. This enables formal specifications to guide control input selection without requiring the synthesis of a~dedicated safety controller.

\section{Conclusion}
\label{sec:conclusion}

We presented \SplitLTL, a temporal logic formalism for runtime assurance that separates reasoning about the observed execution from reasoning about predicted future behaviors. Past-time monitors incrementally evaluate requirements over the observed history, while future-time monitors assess candidate continuations against the requirements activated by that history. This split provides a direct mechanism for filtering predicted behaviors according to formal assurance requirements. Although developed for \LTL, the approach is not tied to a particular temporal-logic formalism. The same architecture can accommodate other formalisms whose specifications admit past- and future-oriented components with efficient monitoring procedures. \SplitLTL therefore provides a general pattern for integrating temporal-logic monitoring with prediction-based decision making.

The performance of monitors becomes particularly critical in such an assurance framework, as modern autonomous systems may generate thousands of candidate behaviors over prediction horizons of up to 30 seconds, with each behavior consisting of hundreds of states. Consequently, evaluating formal requirements over all candidate behaviors within the controller's decision cycle can impose substantial computational overhead. Together with the need to monitor multiple and potentially overlapping specifications, this motivates moving beyond standalone monitoring toward the orchestration of multiple monitors, where monitoring resources can be coordinated and computation shared across predictions and specifications.

Beyond monitoring efficiency, we are also interested in making runtime verification a reusable assurance component for supporting learning-based prediction mechanisms. Runtime verdicts for predicted behaviors can be used as feedback to guide or refine the prediction mechanism in subsequent control cycles. This creates a feedback loop in which formal monitoring can inform future prediction while remaining an independent assurance layer. Regardless of how candidate behaviors are generated, each newly generated behavior is evaluated against the formal requirements using the monitoring procedure presented in this paper. Such an architecture allows runtime verification to support learning-based prediction without requiring the prediction mechanism itself to provide formal guarantees.

Together, these directions position \SplitLTL as a specification-based assurance layer that can operate between prediction, monitoring, and decision making in autonomous systems.

\section*{AI Usage and Disclosure}
\label{sec:disclosure}

Generative AI tools were used to assist with language editing, proofreading, and improving the clarity and organization of this manuscript. The author reviewed and verified all AI-assisted content and remains fully responsible for its accuracy, originality, and integrity.

\bibliographystyle{eptcs}
\bibliography{references}

\end{document}